\documentclass[%
 reprint,
superscriptaddress,
 amsmath,amssymb,
 aps,
]{revtex4-2}

\usepackage{graphicx}
\usepackage{dcolumn}
\usepackage{bm}

\usepackage{setspace}
\usepackage{textcomp}
\usepackage{gensymb}
\usepackage[version=3]{mhchem}
\usepackage{multirow, makecell}
\usepackage{graphicx}

\begin{document}

\preprint{APS/123-QED}

\title{Structural Phase Transition and Superconductivity in 2H-BaGaGe with Buckled Honeycomb Layers}


\author{Dorota~I.~Walicka}
\affiliation{Department of Quantum Matter Physics, University of Geneva, CH-1211 Geneva, Switzerland}
\affiliation{Department of Chemistry, University of Zurich, CH-8057 Zurich, Switzerland}

\author{Robin~Lef\`{e}vre}
\affiliation{Department of Chemistry, University of Zurich, CH-8057 Zurich, Switzerland}

\author{Olivier~Blacque}
\affiliation{Department of Chemistry, University of Zurich, CH-8057 Zurich, Switzerland}

\author{Sara~A.~L\'{o}pez-Paz}
\affiliation{Department of Quantum Matter Physics, University of Geneva, CH-1211 Geneva, Switzerland}

\author{Carl~W.~Rischau}
\affiliation{Department of Quantum Matter Physics, University of Geneva, CH-1211 Geneva, Switzerland}

\author{Antonio~Cervellino}
\affiliation{Laboratory for Synchrotron Radiation - Condensed Matter, Paul Scherrer Institut, CH-5232 Villigen, Switzerland}

\author{Carlos~A.~Triana}
\affiliation{Department of Chemistry, University of Zurich, CH-8057 Zurich, Switzerland}

\author{Fabian~O.~von~Rohr}
\affiliation{Department of Quantum Matter Physics, University of Geneva, CH-1211 Geneva, Switzerland}


\begin{abstract}

We report on the structural and superconducting properties of the intermetallic compound BaGaGe. We find that this material undergoes a structural second-order phase transition from the distorted \ce{AlB2}-type structure (1H, $a$ = 4.3254(2) \r{A}, $c$ = 5.1078(3) \r{A},  \textit{P}6/\textit{mmm}) into the \ce{CaIn2}-type structure (2H, $a$ = 4.3087(3) \r{A}, $c$ = 10.2117(6) \r{A}, \textit{P}6\textsubscript{3}/\textit{mmc}) at a transition temperature of $T_{\rm S}$ = 253 K. We find that the structural phase-transition corresponds to a coherent buckling of the honeycomb layers, which we can interpret as a disorder-to-order transition of the atoms located within this layer. We show that the 2H-BaGaGe phase becomes superconducting at a critical temperature of $T_{\rm c}$ = 2.1 K. The bulk nature of the superconductivity in 2H-BaGaGe is confirmed by means of specific heat measurements, where we determine a value of $\Delta C$/$\gamma T_{\rm c}$ = 1.59, which is close to the expected BCS value in the weak coupling limit.


\end{abstract}

\maketitle

\raggedbottom                              
\maketitle
\sloppy


\begin{figure*}[!]
	\includegraphics[width=1\linewidth]{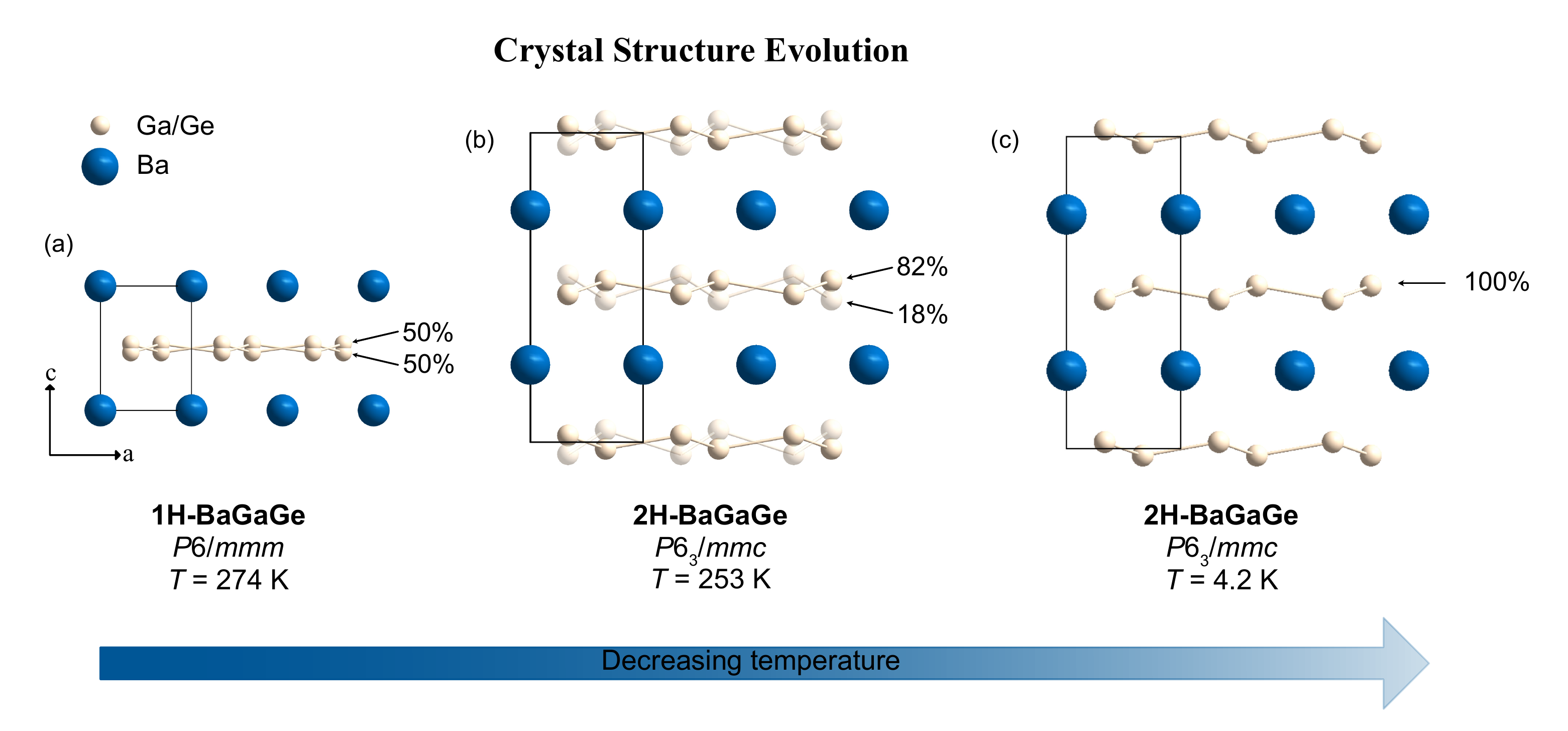}
	\caption{Crystal structure of BaGaGe with varying temperature. (a) Structure from SXRD at 274 K: 1H modification with statistical disorder around the mirror plane perpendicular to the $c$-axis leading to a 50\%/50\% probability of the buckling in one direction of the honeycomb lattice. (b) Structure from SXRD at 253 K: 2H modification with pronounced buckling, with a disorder site occupancy of 82\%/18\% in favor of one direction. (c) Structure from synchrotron PXRD at 4.2 K: fully ordered 2H structure.}
	\label{fig:1}
\end{figure*}

\begin{figure}[h!]
	\includegraphics[width=1\linewidth]{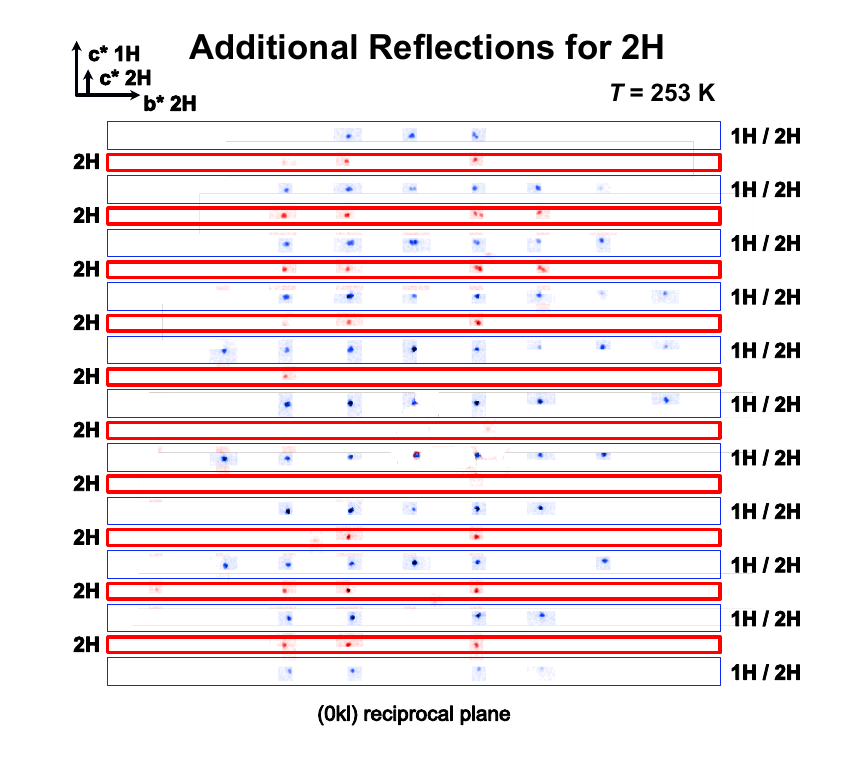}
	\caption{Reconstruction of the reciprocal space along the (0kl) reciprocal plane with additional reflections originating from the 2H-BaGaGe structure.}

	\label{fig:ewald}
\end{figure}


\section{Introduction}\label{sec:introduction}


Superconductivity in the \ce{AlB2}-type compounds with honeycomb layers has been widely studied since the discovery of superconductivity with a critical temperature of $T_{\rm c}$ = 39 K in \ce{MgB2} \cite{Nagamatsu2001}. Prominent examples of the superconductivity on the honeycomb structure include the proposal for intrinsic topological superconductivity in SrPtAs \cite{Nishikubo2011,Biswas2013,Fischer2014}, as well as superconductivity in magic-angle bi-layer graphene \cite{Cao2018}.

There is a wide range of structural derivatives related to the \ce{AlB2} structure with honeycomb layers. The main difference between these structures is the arrangement of the atoms within the honeycomb layers, the stacking sequences of the layers, as well as the buckling of the layers. Especially in ternary compounds, group-subgroup relationships lead to, among others, compounds with ZrBeSi-, \ce{CaIn2}-, and YPtAs-type structures \cite{Hoffmann2001}. Surprisingly, among over the $\approx$ 615 known intermetallic compounds in these structure-types \cite{Dshemuchadse2015}, only a few have been found to be superconducting \cite{Nishikubo2011,Biswas2013,Fischer2014,Lorenz2003,Walicka2021}. 

The compound CaAlSi has the highest critical temperature among all reported ternary \ce{AlB2}-type superconductors, with a $T_{\rm c} = $ 7.8 K \cite{Imai2003}. It crystallizes into the \textit{P}6\textsubscript{3}/\textit{mmc} space group with a so-called 6H structure. This structure contains six hexagonal layers of six-membered-rings of [AlSi]$^{2-}$ units in its unit cell. These honeycomb layers are stacked on top of each other, and all of them are slightly buckled. Recently, it was reported -- by means of specific heat measurements and muon spin rotation spectroscopy -- that the buckling of the layers seems to be correlated to an enhancement of the superconducting critical temperature in the \ce{Ca_{1-x}Sr_xAlSi} solid solution \cite{Walicka2021}.

\ce{MgB2} and the ternary nine-electron 111 systems are both considered to be conventional superconductors with electron-electron pairs mediated by phonons. However, the mechanisms of their electron-phonon coupling are believed to be notably different. In the case of \ce{MgB2}, the superconducting characteristics can be attributed to the partially filled B-B bonding states that exhibit a strong connection with the in-plane boron vibrations \cite{mazin2002superconductivity,choi2002origin}. While for example, in the case of CaAlSi theoretical analyses have pinpointed the partially filled $\pi^*$ combined with an ultrasoft mode linked to the out-of-plane Al-Si vibration, as the potential foundation for the superconductivity \cite{mazin2004electronic}.

Nine-electron 111 compounds with honeycomb layers can be considered as charge imbalanced Zintl phases \cite{Evans2009,Giantomassi2005}. The ideal electronic structure of a planar hexagonal layer of these compounds leads to two $\pi$ bands, one of which is bonding, while the other one is anti-bonding. Buckling of these honeycomb layers destroys the $\pi$ bands and creates lone pairs in the electronic states. While at the same time, the soft mode that is responsible for the buckling of the layers is believed to be connected to the superconductivity of these compounds (see, e.g. reference \cite{mazin2004electronic}). Generally, in nine-electron 111 compounds with honeycomb layers, the superconductivity is the result of a complex interplay among various factors. These include the interrelationship of the structural elements, the associated phonon modes, the ensuing electronic modifications, and the cumulative impact of these aspects on superconductivity. 

Within this family of compounds, BaGaGe, was first reported to crystallize into the YPtAs-type structure \cite{czybulka1989}, however further studies have shown that it crystallizes in the \ce{AlB2}-type, i.e. the 1H structure with planar honeycomb layers \cite{evans2008,Evans2009}. Preliminary data on 1H-BaGaGe suggested a transition to a superconducting state at a critical temperature of $T_{\rm c}$ = 2.4 K \cite{Evans2009}.

In this work, we study the structural and physical properties of BaGaGe. Specifically, we show that BaGaGe crystallizes in a distorted version of the \ce{AlB2}-type structure (1H) with buckled honeycomb layers and a statistical disorder around a mirror plane perpendicular to $c$ axis and that it undergoes a structural phase transition from the 1H to the 2H-type structure at $T_{\rm S}$ $\approx$ 253 K , which corresponds to a coherent buckling in favor of one direction of the honeycomb layers. We find that this 2H-BaGaGe compound then undergoes a transition to a bulk superconducting state at low temperatures, which we characterize by means of resistivity, magnetization, and specific heat measurements, providing a comprehensive overview of the physical properties.


\begin{figure*}
	\includegraphics[width=1\linewidth]{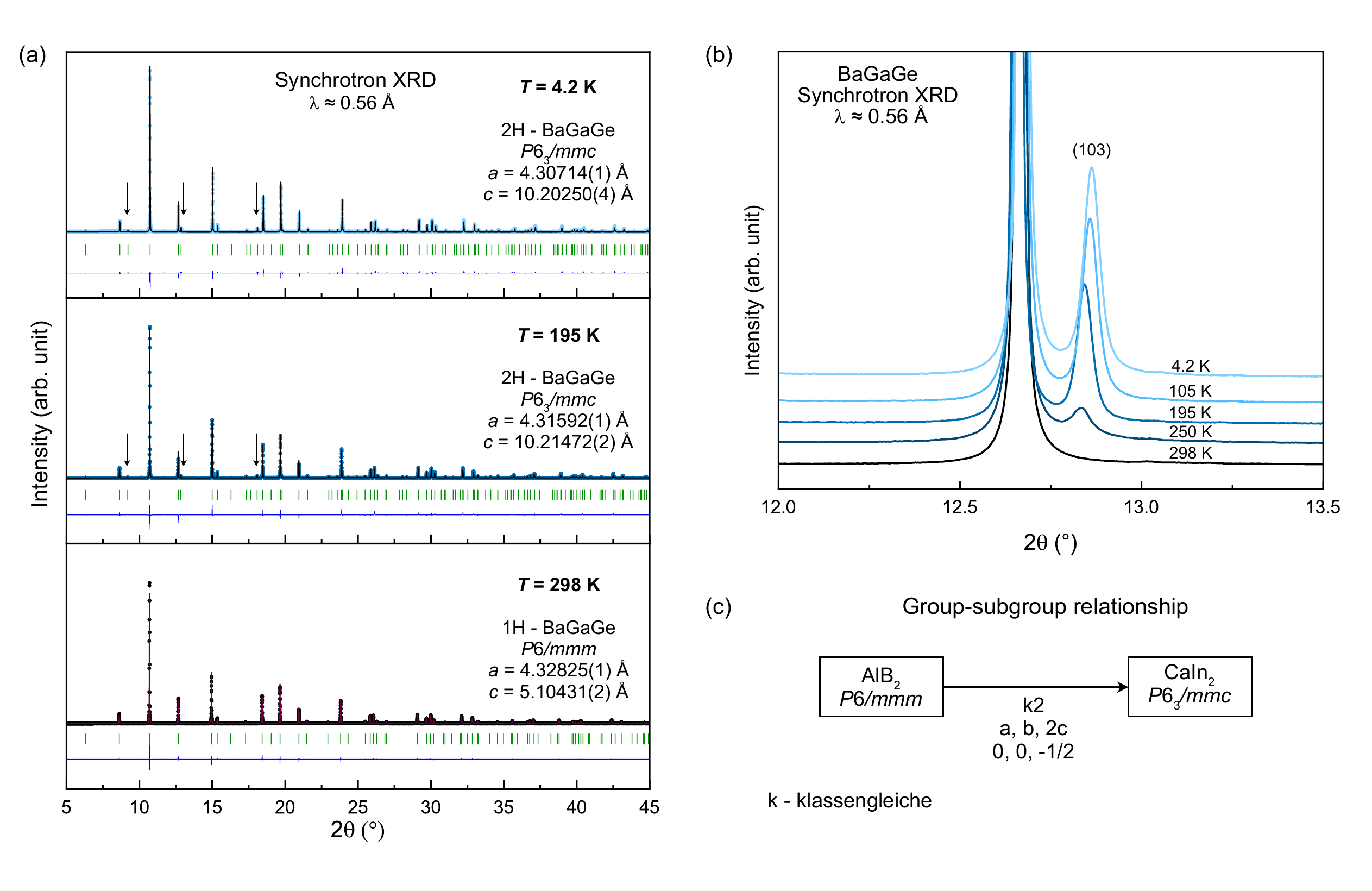}
	\caption{(a) Rietveld refinement of the synchrotron PXRD patterns measured at $T =$ 4.2, 195 and 298 K. The arrows indicate the position of selected additional peaks originating from the 2H structure. (b) Evolution of the (103) peak with the temperature from synchrotron PXRD. (c) Group-subgroup relationship between 1H-BaGaGe and 2H-BaGaGe. Below the arrow the index (k2), unit cell transformation (a, b, 2c) and origin shift (0, 0, -1/2) are indicated.}
	\label{fig:synchrotron}
\end{figure*}

\section{Methods}\label{sec:exp}
\textbf{Synthesis:} BaGaGe was prepared by arc melting under argon of stoichiometric amounts of Ba (Sigma Aldrich, 99.99\%), Ga (Roth, 99.999\%), and Ge (Alfa Aesar, 99.999\%). The arc melting was performed on a water-cooled copper plate using a tungsten tip. The sample was flipped over and completely molten three times, to insure homogeneity, while avoiding the evaporation of the elements. After arc melting the sample was sealed into a quartz tube (9 mm diameter) and annealed at 830 $^\circ$C for 24 h followed by cooling down to room temperature at a rate of 30\degree C/h. 

\textbf{Diffraction:} Temperature dependent single-crystal X-ray diffraction (SXRD) measurements were collected using Cu-K\textsubscript{$\alpha$1} ($\lambda$ = 1.54184 \r{A}) radiation with a XtaLAB Synergy, Dualflex, Pilatus 200 K. The measured temperatures are listed in tables I, II and III in the supplementary material. Pre-experiment, data collection, manual data reduction, and absorption corrections were carried out with the program suite CrysAlisPro \cite{Crysalispro}. Using the Olex2 crystallography software \cite{Dolomanov2009}, the structure was solved using the SHELXT Intrinsic Phasing solution program \cite{Sheldrick2015} and refined with the SHELXL 2018/3 program \cite{Sheldrick2015a} by full-matrix least-squares minimization on F$^2$. 

Temperature-dependent synchrotron powder X-ray diffraction (PXRD) experiments were carried out at the Swiss Light Source (SLS) at the Paul Scherrer Institute (PSI), Villigen, Switzerland, at the Materials Science (MS) X04SA beamline. The wavelength $\lambda$ = 0.563270(6)~\AA~was refined using a \ce{LaB6} standard. Measurements in the temperature range from 4.2 K to 298 K were performed on powdered sample filled in a 0.3 mm glass capillary. Data were analyzed by means of Rietveld refinement using the FullProf program suite \cite{Rodriguez1993}.

\textbf{Elemental analysis:} Elemental characterization was performed using a Zeiss GeminiSEM 450 field emission scanning electron microscope equipped with an X-ray detector X-MAX80, AZTec Advanced, Oxford. A total of ten spectra were collected at an applied voltage of 30 kV and at 12 mm working distance.

\textbf{Physical Properties:} The temperature dependent magnetization measurements were performed using a Quantum Design Magnetic Properties Measurement System (MPMS3) equipped with a vibrating sample magnetometry (VSM) option and a 7 T magnet. Resistivity and specific heat were measured using a Quantum Design Physical Property Measurement System (PPMS) with a Helium-3 option for measurement below 1.8 K. For resistivity measurements the four probe method was applied, where golden wires were connected to the sample with silver paste. Heat capacity measurements in a temperature range between $T$ = 1.8 and 200 K were performed with use of N Apiezon grease, while H Apiezon grease was used for the measurements between $T$ = 200 and 300 K.

\section{Results and discussion}\label{sec:results}

\subsection{Structural Characterization}\label{sec:structure}

Phase-pure silvery shiny samples of BaGaGe were obtained. The samples were found to be brittle, and could be easily broken into small crystallites. The stoichiometry of the synthesized compound was measured by Energy Dispersive X-ray Spectroscopy (EDX) analysis and the average atomic weight were calculated to be: Ba 34.03 $\pm$ 0.38 \%, Ga 32.67 $\pm$ 0.59 \%, Ge 33.30 $\pm$ 0.41 \%, which corresponds to a composition of \ce{Ba_{1.02}Ga_{0.98}Ge_{1.00}}, close to the target 111 ratio. 

The structure of BaGaGe was solved by means of synchrotron PXRD between $T=$ 298 and 4.2 K, and with SXRD in the temperature range between $T=$ 350 and 100 K. The evolution of the structure of BaGaGe is summarized in figure \ref{fig:1}, and we will discuss it in detail in the following. 

Using SXRD the structural phase transition that BaGaGe undergoes from the 1H-type structure to the 2H-type structure could be resolved in detail. The Ge and Ga atomic positions could not be sufficiently distinguished with X-ray diffraction due to the similar scattering factor (close atomic numbers). Hence, in our structural model, a mixed occupancy was used of each site, i.e. 50\% of the total site occupancy by Ge and 50\% by Ga. Previous reports on BaGaGe \cite{Evans2009,evans2008} had found that at room temperature it crystallize into a \ce{AlB2}-type structure where the atoms withing the honeycomb lattice occupy the 2d Wyckoff position x = 2/3, y = 1/3 and z = 1/2. However, our structural model for BaGaGe at high temperature suggests a 1H structure with buckled honeycomb layers, where the Ga/Ge atoms are located at 4h Wyckoff position x = 2/3, y = 1/3 and z = 0.4647.

\begin{figure}
	\includegraphics[width=1\linewidth]{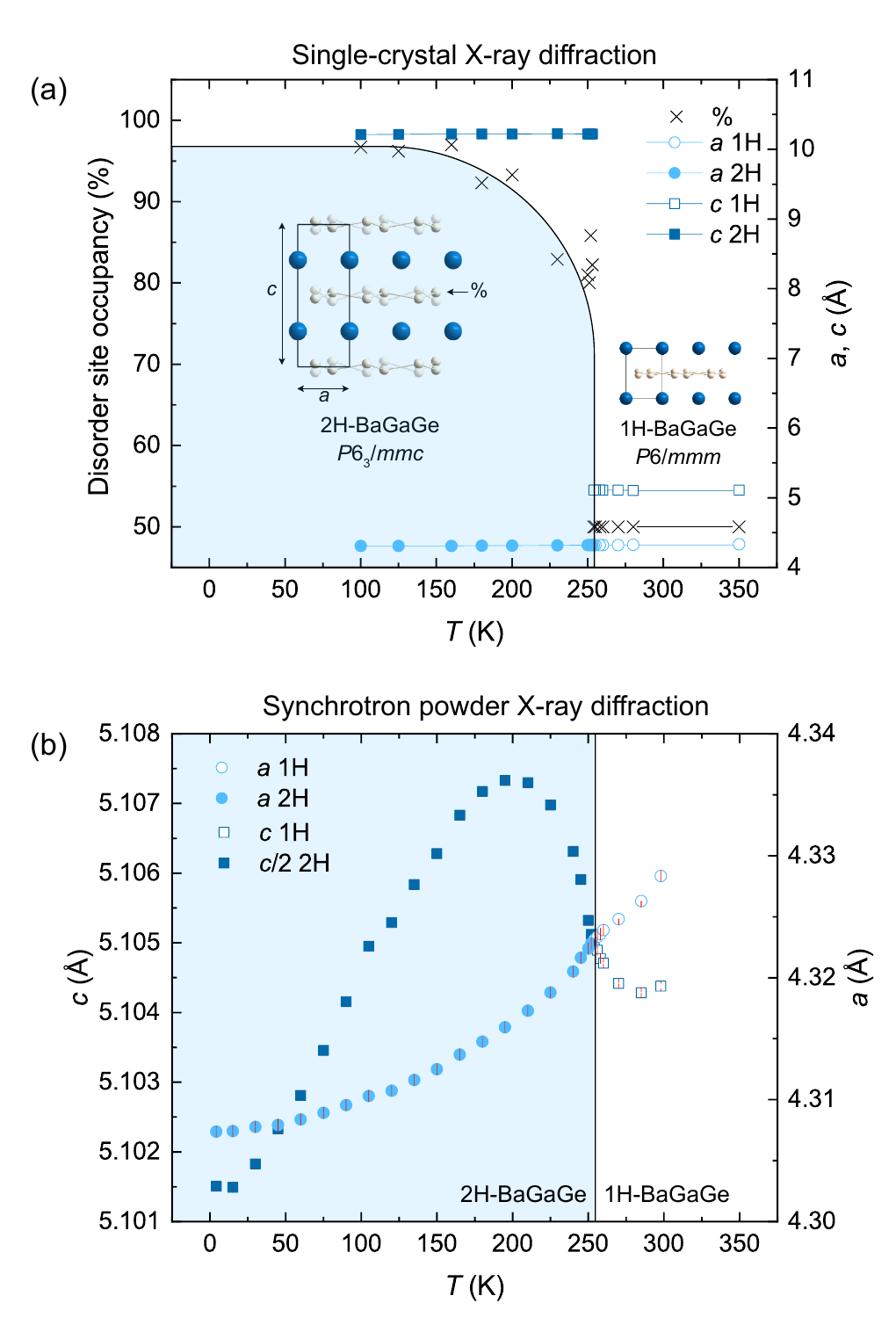}
	\caption{(a) Unit cell parameters and disorder site occupancies from SXRD data. (b) Evolution of the unit cell parameters with temperature obtained from synchrotron PXRD. The $c$-axis for the 2H-BaGaGe was dived by 2 for better comparability.}
	\label{fig:unitcell}
\end{figure}

Four representative refinements from SXRD data measured at temperatures of $T$ = 274, 255, 253, and 99.9 K are summarized in table \ref{table:singlecrystal} and the corresponding atomic positions are reported in table \ref{table:atoms}. The data of all the other measurements are provided in the SI. From 350 K down to 255 K the 1H structure was resolved with a statistical disorder around the mirror plane perpendicular to the $c$-axis, leading to a 50\%/50\% probability for the buckling of the Ga/Ge layers in one direction, as shown in figure \ref{fig:1}(a). Below a temperature of $T$ = 255 K we observe the structural phase transition, in which the buckling of the atoms is favored to face each other perpendicular to the $c$-axis, which lowers the symmetry and causes doubling of the unit cell along the $c$-axis. The refinement at a temperature of $T$ = 254 K did not lead to a unique result for any of the two structures. However, the refinement at a so-called transition temperature $T_{\rm S}$ = 253 K clearly shows that the structure crystallizes in the \textit{P}6\textsubscript{3}/\textit{mmc} space group, with the full structural solution depicted in figure \ref{fig:1}(b). At this $T_{\rm S} =$ 253 K the occupancy jumps from a 50\%/50\% probability of random order to a 82\%/18\% favored order. At the lowest recorded temperature using SXRD, i.e. \textit{T} = 100 K, a 97\% ordering of the buckling is found. Figure \ref{fig:ewald} shows the reconstruction of the reciprocal space along the (0kl) reciprocal plane, with the additional reflections originating from the 2H structure. Refined unit cell parameters together with disorder site occupancies across the structural phase transition obtained from SXRD are summarized in figure \ref{fig:unitcell}(a). 

To confirm the results obtained from SXRD, we performed synchrotron PXRD experiments. The differences between 1H and 2H structures are subtle, though clearly observable in the high-resolution of the synchrotron PXRD data. Hence, the Bragg reflections originating from the 2H structure could be conclusively resolved. In figure \ref{fig:synchrotron}(a), we show three representative synchrotron PXRD pattern collected at temperatures of $T =$ 4.2, 195 and 298 K, along with the respective Rietveld refinements. Figures 1 and 2 in the SI show the full set of synchrotron PXRD patterns. The cell parameters and agreement factors are summarized in table IV in the SI. The Rietveld refinement of the synchrotron PXRD pattern collected at $T =$ 298 K (see bottom figure \ref{fig:synchrotron}(a)) confirmed that BaGaGe crystallizes in the \textit{P}6/\textit{mmm} space group. The 1H-BaGaGe structure remains the stable configuration for the temperature range from $T$ = 298 to 256 K. Below $T_{\rm S}$, the synchrotron PXRD patterns in the temperature range between $T$ = 254 and 4.2 K display additional (k0l) and (hkl) diffraction peaks, namely the (101), (105), (203), (205), (116), (213), (215), (101), (207), (313), (209), (402), (1011), (407), (317), and (1013) reflections. For clarity, the additional Bragg reflections with the highest intensities -- originating from the 2H structure -- are highlighted with arrows in the middle and top panels of figure \ref{fig:synchrotron}(a). The intensity of these reflections progressively increases as we observe the disorder-to-order transition within the honeycomb lattice. The example of the intensity evolution of the (103) Bragg reflection is shown in figure \ref{fig:synchrotron}(b). At base temperature, i.e. $T =$ 4.2 K the 2H structure is fully ordered. For visualization purposes, the structure of 2H-BaGaGe at a temperature of $T =$ 4.2 K is depicted in the figure \ref{fig:1}(c).

The \ce{AlB2} (1H) and \ce{CaIn2} (2H) structure types are connected through a group-subgroup relation tree by a klassengleichen transformation \cite{Hoffmann2001}. This can be understood that both structures belong to the same class, but the unit cell has to be enlarged for the 2H structure, which also reduces other symmetry elements, i.e. the 6-fold rotation axis is replaced with a 6\textsubscript{3} screw axis and the mirror plane $m$ is replaced by a $c$ axial glide plane. The unit cell transformation and shift of the origin are shown in figure \ref{fig:synchrotron}(c). The phase transitions between space groups connected by the group-subgroup relationships are usually of a second order \cite{muller2006}, with a progressive conversion over a certain temperature range. Here, in the case of BaGaGe, this can be very well illustrated when observing the evolution of unit cell parameters within one layer as depicted in figure \ref{fig:unitcell}(b). We find the cell parameter \textit{a} to decrease over the whole temperature range with decreasing temperature as expected by the thermal expansion. The cell parameter \textit{c} has a more complex change as it is featuring the structural phase transition. The cell parameter \textit{c} first increases with decreasing temperature due to the progressive buckling of the honeycomb layers, which increases the interlayer distance. Then -- below $T \approx$ 180 K -- the cell parameter \textit{c} decreases with decreasing temperature as expected. A slow, progressive change of the cell parameter \textit{c} over a large temperature range confirms the second-order nature of the 1H to 2H phase transition.

\begin{table*}
\caption{Crystallographic data for single-crystal X-ray diffraction of \ce{BaGaGe} at four different temperatures, with a transition observed around 254 K from \textit{P}6/\textit{mmm} to \textit{P}6\textsubscript{3}/\textit{mmc} space group.}
\label{table:singlecrystal}
\begin{singlespace}
    \centering
    \begin{tabular}{lllll}
        \hline
        \hline
 		\multicolumn{5}{l}{\textbf{Physical, crystallographic, and analytical data}} \\
 		\hline
Formula & 1H-BaGaGe & 1H-BaGaGe & 2H-BaGaGe & 2H-BaGaGe \\
CCDC/FIZ & CSD-2263587 & CSD-2263597 & CSD-2263583 & CSD-2263586 \\
Structure type & \ce{AlB2} & \ce{AlB2} & \ce{CaIn2} & \ce{CaIn2} \\
Mol. wt. (g mol\textsuperscript{-1}) & 279.65 & 279.65 & 279.65 & 279.65 \\
Crystal system & Hexagonal & Hexagonal & Hexagonal & Hexagonal \\
Space group & \textit{P}6/\textit{mmm} & \textit{P}6/\textit{mmm} & \textit{P}6\textsubscript{3}/\textit{mmc} & \textit{P}6\textsubscript{3}/\textit{mmc} \\
Space group number & 191 & 191 & 194 & 194 \\
a (\r{A}) & 4.3254(2) & 4.32097(18) & 4.3204(2) & 4.3087(3) \\
b (\r{A}) & 4.3254(2) & 4.32097(18) & 4.3204(2) & 4.3087(3) \\
c (\r{A}) & 5.1078(2) & 5.1093(2) & 10.2199(5) & 10.2117(6) \\
V (\r{A}\textsuperscript{3}) & 82.759(8) & 82.615(8) & 165.206(17) & 164.18(2) \\
Z & 1 & 1 & 2 & 2 \\
Calculated density (g cm\textsuperscript{-3}) & 5.611 & 5.621 & 5.622 & 5.657 \\[2ex]
\multicolumn{5}{l}{\textbf{Data Collection}} \\
Temperature (K) & 274(6) & 255.00(10) & 253.00(10) & 99.9(2) \\
Radiation & Cu-K$\alpha_1$ & Cu-K$\alpha_1$ & Cu-K$\alpha_1$ & Cu-K$\alpha_1$  \\
 & (1.54184 \r{A}) & (1.54184 \r{A}) &  (1.54184 \r{A}) & (1.54184 \r{A}) \\
Crystal color & \multicolumn{4}{c}{Gray shiny} \\
Crystal size (mm$^{3}$) & 0.02 x 0.02 x 0.01 & 0.02 x 0.01 x 0.01 & 0.02 x 0.02 x 0.01 & 0.02 x 0.01 x 0.01 \\
Linear absorption coefficient (cm$^{-1}$) & 1092.08 & 1093.99 & 1094.15 & 1100.98 \\
Scan mode & $\omega$ & $\omega$ & $\omega$ & $\omega$ \\
Recording range $\theta$ (\degree) & 8.681 --- 73.059 & 11.844 --- 77.012 & 8.677 --- 66.656 & 8.684 --- 66.879 \\
h range & -4 --- +5 & -5 --- +3 & -5 --- +4 & -4 --- +4 \\
k range & -5 --- +5 & -3 --- +5 & -5 --- +4 & -5 --- +3 \\
l range & -5 --- +6 & -6 --- +6 & -9 --- +12 & -10 --- +12 \\
Nb. of measured reflections & 463 & 472 & 825 & 812 \\[2ex]
\multicolumn{5}{l}{\textbf{Data reduction}} \\
Completeness & 100 & 100 & 100 & 100 \\
No. of independent reflections & 55 & 55 & 76 & 76 \\
Rint (\%) & 2.38 & 2.46 & 2.89 & 3.56 \\
Absorption corrections & sphere & sphere & sphere & sphere \\
Empirical corrections & \multicolumn{4}{c}{Frame scaling and spherical harmonics} \\
Independent reflections with I $\geq$ 2.0$\sigma$ & 55 & 55 & 69 & 72 \\[2ex]
\multicolumn{5}{l}{\textbf{Refinement}} \\
R1 (obs / all) (\%) & 1.24/1.24 & 0.94/0.94 & 0.94/1.10 & 0.98/1.23 \\
wR2 (obs / all) (\%) & 2.61/2.61 & 2.27/2.27 & 1.83/1.85 & 2.33/2.39 \\
GOF & 1.310 & 1.215 & 1.079 & 1.185 \\
No. of refined parameters & 7 & 7 & 8 & 8 \\
Difference Fourier residues (eÅ\textsuperscript{-3}) & -0.354 / +0.332 & -0.370 / +0.455 & -0.295/0.247 & -0.259 / 0.477\\
\hline
\hline
    \end{tabular}
    \end{singlespace}
\end{table*}

\begin{table*}
 \caption{Refined coordinates, iso- and anisotropic displacement parameters (ADPs), and their estimated standard deviations for temperature study of the single-crystal X-ray diffraction of \ce{BaGaGe}.}
\label{table:atoms}
\centering
\begin{tabular}{llllllllllll}
\hline
\hline
\textbf{T (K)} & \textbf{Space group} & \textbf{Atom} & \textbf{Wyckoff} & \textbf{Occ.} & \textbf{x} & \textbf{y} & \textbf{z} & \textbf{$U_{iso}$} & \textbf{$U_{11}$} & \textbf{$U_{33}$} & \textbf{$U_{12}$} \\
\hline
\multirow{2}{*}{\textbf{274}} & \multirow{2}{*}{\textit{P}6/\textit{mmm}} & Ba & 1a & 1 & 0 & 0 & 0 & 0.0228(3) & 0.0231(3) & 0.0223(4) & 0.01153(17) \\
& & (Ga/Ge) & 4h & 0.50 & 2/3 & 1/3 & 0.4647(13) & 0.291(11) & 0.0211(5) & 0.045(3) & 0.0106(3) \\[2ex]
\multirow{2}{*}{\textbf{255}} & \multirow{2}{*}{\textit{P}6/\textit{mmm}} & Ba & 1a & 1 & 0 & 0 & 0 & 0.0147(3) & 0.0150(3) & 0.0140(4) & 0.00750(14) \\
& & (Ga/Ge) & 4h & 0.50 & 2/3 & 1/3 & 0.4619(8) & 0.0196(8) & 0.0129(4) & 0.033(3) & 0.00646(18) \\[2ex]
\multirow{3}{*}{\textbf{253}} & \multirow{3}{*}{\textit{P}6\textsubscript{3}/\textit{mmc}} & Ba & 2b & 1 & 0 & 0 & 3/4 & 0.0132(2) & 0.0135(2) & 0.0125(3) & 0.00676(12) \\
& & (Ga/Ge)1 & 4f & 0.82 & 1/3 & 2/3 & 0.48145(8) & \multirow{2}{*}{0.0185(3)} & \multirow{2}{*}{0.0117(3)} & \multirow{2}{*}{0.0322(7)} & \multirow{2}{*}{0.00583(14)} \\
& & (Ga/Ge)2 & 4f & 0.18 & 1/3 & 2/3 & 0.5219(6) &  &  &  &  \\[2ex]
\multirow{3}{*}{\textbf{99.9}} &\multirow{3}{*}{\textit{P}6\textsubscript{3}/\textit{mmc}} & Ba & 2b & 1 & 0 & 0 & 3/4 & 0.0120(3) & 0.0123(3) & 0.0115(4) & 0.00617(15) \\
& & (Ga/Ge)1 & 4f & 0.97 & 1/3 & 2/3 & 0.47832(4) & \multirow{2}{*}{0.0157(3)} & \multirow{2}{*}{0.0117(4)} & \multirow{2}{*}{0.0237(5)} & \multirow{2}{*}{0.00586(18)} \\
 && (Ga/Ge)2 & 4f & 0.03 & 1/3 & 2/3 & 0.5237(15) &  &  &  & \\
 \hline
\multicolumn{11}{l}{Ga/Ge ratio always fixed to 50/50 on each shared position as X-ray do not allow to distinguish Ga/Ge atoms.}\\
\multicolumn{11}{l}{$U_{11}$ = $U_{22}$ = $U_{13}$, $U_{23}$ = 0 due to symmetry constraints.}
\\
\hline
\hline
\end{tabular}
\end{table*}


\subsection{Physical Properties}\label{sec:phys}

\begin{figure}
	\includegraphics[width=1\linewidth]{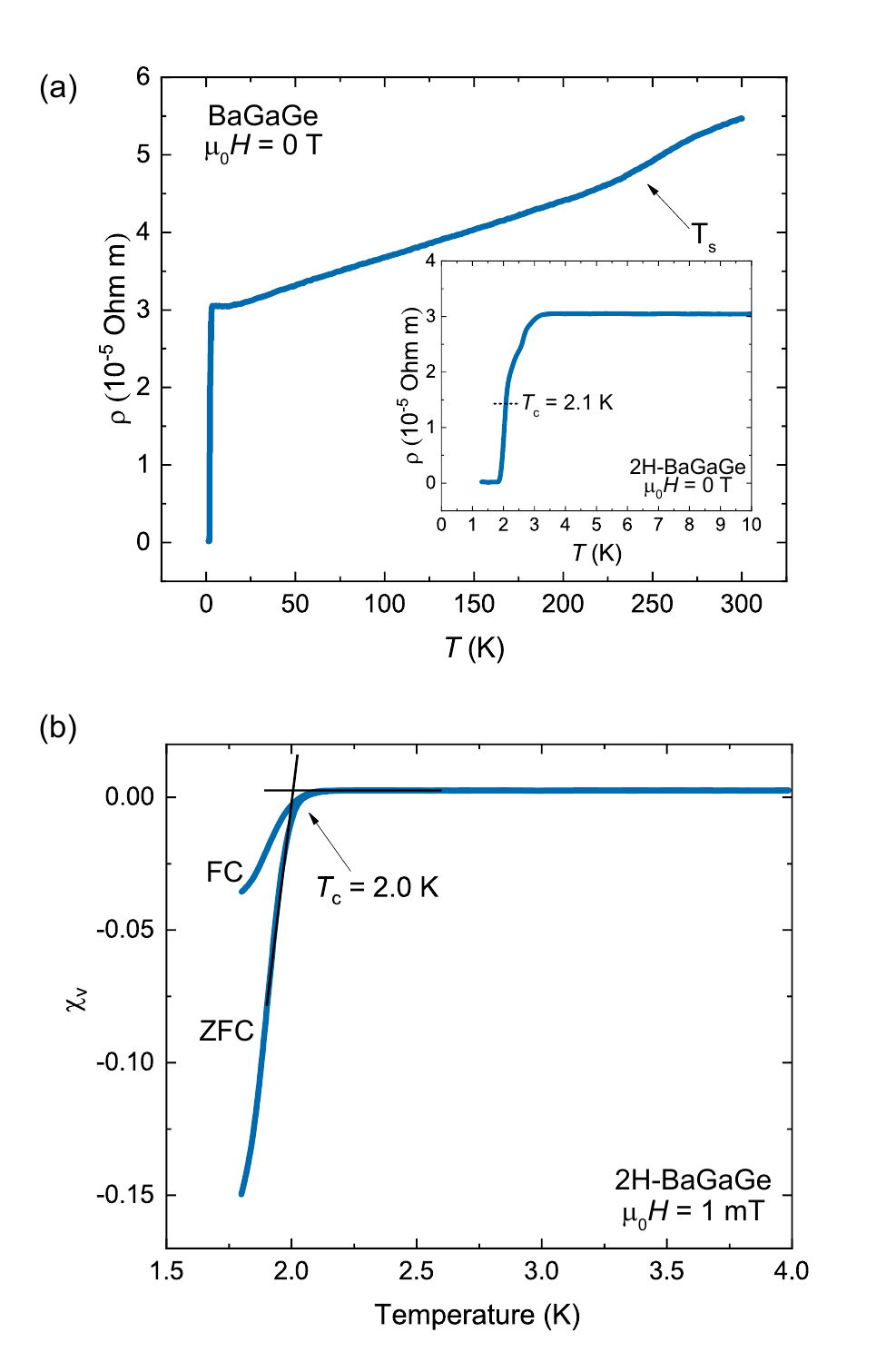}
\caption{(a) Resistivity data in the temperature range between $T$ = 1.3 and 300 K. The inset shows the transition to the superconducting state with $T_{\rm c}$ = 2.1 K assigned at the 50\% drop of $\rho(T)$. (b)  Zero field cooling (ZFC) and filed cooling (FC) magnetization in the temperature range between $T$ = 1.8 and 4 K in an external magnetic field of $\mu_0 H$ = 1 mT, with the $T_{\rm c}$ = 2.0 K in the magnetization.}
	\label{fig:mag-res}
\end{figure}

The temperature-dependent resistivity of BaGaGe in the temperature range between $T$ = 300 and 1.8 K in zero magnetic field $\mu_0 H = 0$ T is depicted in figure \ref{fig:mag-res}(a). The resistivity of BaGaGe decreases with decreasing temperature, displaying a metallic behavior. At the structural phase transition $T_{\rm S}$ = 253 K a clearly pronounced discontinuity is observed, which is highlighted with an arrow in figure \ref{fig:mag-res}(a). We find a residual resistivity ratio (RRR) for the 2H-BaGaGe, here defined as RRR = $\rho$(225 K)/$\rho$(3 K) = 1.54, which indicates a short mean-free path of the charge carriers, i.e., a bad-metal behavior. This value for the RRR is likely associated with the polycrystalline nature of the sample that was measured, and henceforth with the significant contribution of the grain boundaries to the sample's resistivity. At low temperatures, we observe a sharp drop in the resistivity at the transition to the superconducting state. We determine the critical temperature to be $T_{\rm c}$ = 2.1 K at a 50\% decrease of $\rho(T)$ from the resistivity (figure \ref{fig:mag-res}(a) inset).

In the temperature-dependent magnetization measurements, we find that BaGaGe is a Pauli paramagnet in the normal state, with a nearly temperature-independent positive magnetization between $T$ = 300 and 5 K (see figure 3 in the SI) in magnetic fields of $\mu_0 H$ = 0.1 and 1 T. 

In order to characterize the superconducting state by magnetization, we have performed zero field cooled (ZFC) and field cooled (FC) measurements in an external field of $\mu_0 H$ = 1 mT, as shown in figure \ref{fig:mag-res}(b). The superconducting transition is clearly pronounced, with a strong diamagnetic shielding fraction. We determine the critical temperature to be $T_{\rm c}$ = 2.0 K in the magnetization.

\begin{figure}
	\includegraphics[width=1\linewidth]{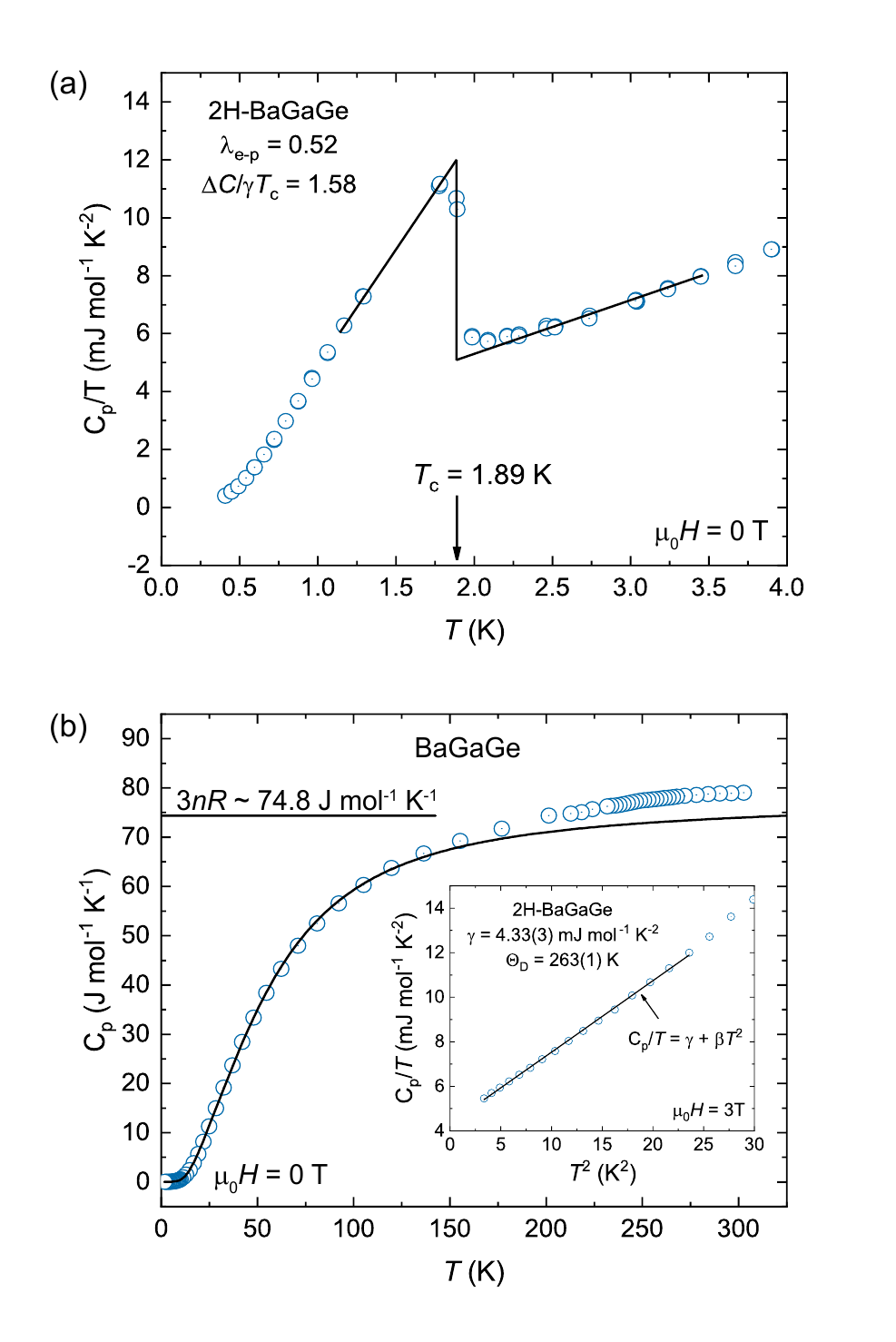}
\caption{(a) Temperature-dependent heat capacity of 2H-BaGaGe in the vicinity of the superconducting phase-transition with an entropy-conserving construction. (b) Heat capacity data in the temperature range between $T$ = 300 K and 2 K. Inset: heat capacity at low temperatures measured in a magnetic field of 3 T.}
	\label{fig:cp}
\end{figure}

In figure \ref{fig:cp}(a,b), we show the temperature-dependent heat capacity of BaGaGe in the superconducting state and in the normal state, respectively. The heat capacity data in the low temperatures (see figure \ref{fig:cp}(b) inset) were analyzed using the equation:

\begin{equation}
    C\textsubscript{p} = C\textsubscript{el} + C\textsubscript{Debye},
    \label{eq:cp0}
\end{equation}

\begin{equation}
C\textsubscript{p}/T = {\gamma}+{\beta}T^{2},
\label{eq:cp1}
\end{equation}

which allowed us to determine $\gamma$ = 4.33(3) mJ mol$^{-1}$ K$^{-2}$ and $\beta$ = 0.321(2) mJ mol$^{-1}$ K$^{-1}$ for 2H-BaGaGe. The Debye temperature, i.e., the temperature of the highest normal mode of vibration, was derived from the $\beta$ coefficient according to: 

\begin{equation}
  \Theta\textsubscript{D} = \Big(\frac{12\pi^{4}}{5\beta} nR\Big)^{1/3},
  \label{eq:cp2}
\end{equation}

where $n$ = 3 and $R$ = 8.31 J mol$^{-1}$ K$^{-1}$ is the gas constant. We obtain a Debye temperature of $\Theta_D$ = 263(1) K for 2H-BaGaGe. 

The jump of the heat capacity corresponds to the transition to the superconducting state and arises from the difference of the specific heat of the electrons in the normal state and when combined in Cooper pairs. With the entropy conserving construction shown in figure \ref{fig:cp}(a) we find a critical temperature in the specific heat of $T_{\rm c}$ = 1.89 K. This value is -- as expected -- slightly lower than the critical temperatures in the resistivity and the magnetization measurements, and therefore in very good agreement with them. The value of $\Delta C$/$\gamma T_{\rm c}$ = 1.59 is slightly larger than the value expected for the weak-coupling BCS theory of 1.43. The value is evidence for 2H-BaGaGe to be a bulk superconductor. 

The obtained values for $T_{\rm c}$ and $\Theta_D$ allow us to estimate the electron-phonon coupling constant $\lambda_{e-p}$, for which we use the semi-empirical McMillan formula \cite{mcmillan1968}:

\begin{equation}
    \lambda_{e-p}=\frac{1.04+\mu^* ln(\Theta\textsubscript{D}/1.45T\textsubscript{c})}{(1-0.62\mu^*)ln(\Theta\textsubscript{D}/1.45T\textsubscript{c})-1.04}.
\label{eq:cp3}
\end{equation}

Here, $\mu^*$ is the Coulomb pseudo-potential parameter, and for our estimation, we have set $\mu^*$ = 0.13. This is an average value used commonly for intermetallic superconductors (see, e.g., references \cite{von2016effect, Walicka2021}). We find an electron-phonon coupling constant of $\lambda_{e-p}$ = 0.52. This value is in good agreement, with 2H-BaGaGe being a superconductor in a weak coupling limit. Furthermore, we find the electron density at the Fermi level from the Sommerfeld coefficient $\gamma$ to be \textit{D(E\textsubscript{F})} = 1.21 states eV$^{-1}$, by applying

\begin{equation}
  D(E\textsubscript{F}) = \frac{3\gamma}{\pi^2 k\textsubscript{B}^2 (1 + \lambda_{e-p})}, 
  \label{eq:cp4}
\end{equation}

where k\textsubscript{B} =  1.38 × 10$^{-23}$ J K$^{-1}$ is the Boltzmann constant.

The heat-capacity data in the normal-state of 2H-BaGaGe -- i.e. below the structural phase transition -- in a temperature range between $T$ =  200 K and 3 K was analyzed using an extension of equation \ref{eq:cp0}: 

\begin{equation}
    C\textsubscript{p} = C\textsubscript{el} + C\textsubscript{Debye} + C\textsubscript{Einstein},
    \label{eq:D-E}
\end{equation}

where the total specific heat is a mixture of the electron (C\textsubscript{el}) and phonon contribution (C\textsubscript{Debye} and C\textsubscript{Einstein}): 

\begin{equation}
    C\textsubscript{el} (T) = \gamma T,
    \label{eq:cel}
\end{equation}

\begin{equation}
    C\textsubscript{Debye} (T) = 9nRk \left(\frac{T}{\Theta_D}\right)^3 \int_0^{\Theta_D} \frac{x^4e^x}{(e^x -1)^2} \ dx,  \
    \label{eq:D}
\end{equation}

\begin{equation}
     C\textsubscript{Einstein} (T) = 3nR(1-k) \left(\frac{\Theta_E}{T}\right)^2 \frac{e^{\Theta_E / T}}{(e^{\Theta_E / T} - 1)^2}.
     \label{eq:E}
\end{equation}

Here, $\Theta_{\rm E}$ is the Einstein temperature, $k$ is a weight factor and $\gamma$ = 4.33(3) mJ mol$^{-1}$ K$^{-2}$, obtained from the low-temperature fit, was used. We obtain $\Theta_{\rm D}$ = 257(9) K,  $\Theta_{\rm E}$ = 91(6) K and k = 0.73. The fitting of the equations \ref{eq:D-E} is shown in figure \ref{fig:cp}(b) as a black line. Above 100 K the data starts to diverge from the fit due to the broad structural phase transition. The value of $\Theta_D$ is in very good agreement with the value obtained from the low temperature data, as well as value of the $\Theta_E$ additionally estimated to be 96 K (see figure 4 in the SI). In agreement with the Dulong-Petit law $C_{\rm p}(T)$ data at the high-temperature approach the value of $3nR$ = 74.8 J mol$^{-1}$ K$^{-1}$, where $n$ is the number of atoms per unit cell and $R$ = 8.31 J mol$^{-1}$ K$^{-1}$ is the gas constant.

\section{Summary and Conclusion}
\label{sec:conclusion}

In this work, we have analyzed the structural and superconducting properties of BaGaGe. We have shown an improved model of the BaGaGe structure at high temperatures, where BaGaGe crystallizes in a distorted version of the \ce{AlB2}-type structure with buckled honeycomb layers and a statistical disorder around a mirror plane. Moreover, the compound undergoes a second-order structural phase transition from the 1H-BaGaGe structure to the 2H-BaGaGe structure with coherent buckling in favor of one direction, which we interpret as an disorder-to-order phase transition. We have investigated the details of the temperature evolution of the buckling of the honeycomb Ga/Ge layer by means of SXRD and synchrotron PXRD. We showed that the structural phase transition happens over an extended temperature range with a maximum at $T_{\rm S}$ = 253 K. The structural phase transition can also be understood as a doubling of the unit cell along the $c$-axis as a consequence of a disorder-to-order phase transition in the honeycomb layer. By means of SXRD we have concluded that in the temperature range from 298 to 254 K the system displays 50\%/50\% positional disorder of Ga/Ge atoms while below 253 K the honeycomb layer starts to buckle in favor of one direction. At 253 K, we have observed a 82\%/18\% ratio of the electron density in favor of one direction, which progressively orders towards the fully buckled structure below 100 K. The synchrotron PXRD allowed us to confirm the presence of the 2H structure down to 4.2 K. We measured resistivity, magnetization, and specific heat in order to fully characterize the normal-state as well as the superconducting properties of 2H-BaGaGe. All three measurements show that BaGaGe undergoes an electronic transition to a bulk superconducting state with a critical temperature of $T_{\rm c}$ = 2.0 K. Furthermore, from the specific heat we have obtained a $\Delta C$/$\gamma T_{\rm c}$ = 1.59, which is close to the expected BCS value and is evidence for the bulk nature of the superconductivity in 2H-BaGaGe in the weak coupling limit. We find BaGaGe to be a versatile honeycomb compound that displays superconductivity in the vicinity of a structural phase boundary.

\section{Acknowledgements}
\label{sec:acknowledgements}

Energy-dispersive X-ray spectroscopy experiments were performed with equipment maintained by the Center for Microscopy and Image Analysis at the University of Zurich. This  work  was  supported by the Swiss National Science Foundation under Grant No. PCEFP2\_194183.


\bibliography{bib}

\end{document}